\documentclass[%
 aip,
 jmp,%
 amsmath,amssymb,
reprint,%
]{revtex4-1}

\usepackage{graphicx}% Include figure files
\usepackage{dcolumn}% Align table columns on decimal point
\usepackage{bm}% bold math
\usepackage{hyperref}
\usepackage{natbib}
\usepackage{epstopdf}
\begin{document}

%\preprint{AIP/123-QED}

\title{A sentiment-based modeling and analysis of stock price during the COVID-19: U- and Swoosh-shaped recovery}% Force line breaks with \\
%\thanks{Footnote to title of article.}

\author{Anish Rai}
\email{anishrai412@gmail.com}
\affiliation{Department of Physics, National Institute of Technology Sikkim, Sikkim, India-737139.}

\author{Ajit Mahata}
\email{ajitnonlinear@gmail.com}
\affiliation{Department of Physics, National Institute of Technology Sikkim, Sikkim, India-737139.}

\author{Md.Nurujjaman}
\email{md.nurujjaman@nitsikkim.ac.in}
\affiliation{Department of Physics, National Institute of Technology Sikkim, Sikkim, India-737139.}
	
\author{Sushovan Majhi}
\email{smajhi@berkeley.edu}
 \affiliation{School of Information, University of California, Berkeley, USA }

\author{Kanish Debnath}
\email{kanish.debnath@flame.edu.in}
 \affiliation{Department of Economics, Flame University, Pune, India-412115 }

\date{\today}% It is always \today, today,

\begin{abstract}
Recently, a stock price model is proposed by A. Mahata et al. [Physica A, 574, 126008 (2021)] to understand the effect of COVID-19 on stock market. It describes V- and L-shaped recovery of the stocks and indices, but fails to simulate the U- and Swoosh-shaped recovery that arises due to sharp crisis and prolong drop followed by quick recovery (U-shaped) or slow recovery for longer period (Swoosh-shaped recovery). We propose a modified model by introducing a new variable $\theta$ that quantifies the sentiment of the investors. $\theta=+1,~0,~-1$ for positive, neutral and negative sentiment, respectively. The model explains the movement of sectoral indices with positive $\phi$ showing U- and Swoosh-shaped recovery. The simulation using synthetic fund-flow ($\Psi_{st}$) with different shock lengths ($T_S$), $\phi$, negative sentiment period ($T_N$) and portion of fund-flow ($\lambda$) during recovery period show U- and Swoosh-shaped recovery. The results show that the recovery of the indices with positive $\phi$ becomes very weak with the extended $T_S$ and $T_N$. The stocks with higher $\phi$ and $\lambda$ recover quickly. The simulation of the Nifty Bank, Nifty Financial and Nifty Realty show U-shaped recovery and Nifty IT shows Swoosh-shaped recovery. The simulation result is consistent with the real stock price movement. The time-scale ($\tau$) of the shock and recovery of these indices during the COVID-19 are consistent with the time duration of the change of negative sentiment from the onset of the COVID-19. This study may help the investors to plan their investment during different crises.
%Valid PACS numbers may be entered using the \verb+\pacs{#1}+ command.
\end{abstract}

%\pacs{05.45.Tp, 89.65.Gh}% PACS, the Physics and Astronomy
%\keywords{Stock market crash, COVID-19 pandemic, Investors Sentiment, Time series, Statistical significant test, stock price model}%Use showkeys class option if keyword
\maketitle

\begin{quotation} 
During the COVID-19 pandemic, the stock market experienced a massive crash in March 2020. All the indices and stocks witnessed a major fall, but their recoveries were not the same. Different types of recoveries are seen during the pandemic, such as V-, L-, U-, Swoosh-shaped recovery. The dynamics of stock market crashes can be investigated with the help of mathematical models. An existing model explains the V- and L-shaped recovery of stocks and indices during the COVID-19 panidemic. However, the model have limitations in explaining the U- and Swoosh-shaped recoveries. We have modified the existing model to understand the U- and Swoosh-shaped recovery of the stock price. The main parameter is the financial antifragility of the company. Fund-flow by the institutional investors and sentiment of the investors are the variables used in the model. The model simulation of the stock price during COVID like shock has been studied using different shock lengths, negative sentiment lengths, financial antifragility and fund-flow. The model simulation is also carried out to study the U- and Swoosh-shaped recovery of Indian sectoral indices during the COVID-19 pandemic. The time-scales of these sectors are calculated using the Hilbert-Huang transform. The analysis may help the investors plan their entry and exit positions during such crisis.
\end{quotation}

\section{\label{sec:intro}Introduction}
Modeling stock price dynamics is an important objective for the stock market, and a reasonably accurate prediction can result in a good profit.~\cite{ballings2015evaluating} Various models, algorithms and data mining techniques are often developed and applied to predict the behaviour of stock prices.~\cite{ballings2015evaluating,naimzada2015financial,naimzada2015introducing,barak2015developing,ou2009prediction,oh2002analyzing,huang2005forecasting,kim2003financial} While models have been developed with parameters like return and dividend~\cite{ding1993long,marsh1987dividend,marsh1986dividend}, others that adjust for seasonal stock trading are also available to create a profitable trading strategy.~\cite{booth2014automated} Due to the complex nature of the stock markets, modeling the complete dynamics is very difficult.~\cite{mantegna1999introduction,huang2003applications,nurujjaman2020time,sornette2004stock,mahata2020identification} However, modeling the stock market dynamics to study the effect of a crisis like the COVID-19 may be possible and useful.~\cite{charles2014large}

The COVID-19 pandemic shook the stock market around the globe severely.~\cite{ashraf2020economic,topcu2020impact,baker2020unprecedented,zhang2020financial,bbc,rai2021statistical} The panic among the investors due to the pandemic was so severe that the market crashed beyond justification within few days.~\cite{rai2021statistical,mahata2021characteristics,mazur2021covid} During the COVID-19 crisis, there were many quality stocks and sectors, which were resilient to the crisis and outperformed the overall market.~\cite{mahata2021modeling,mazur2021covid,asness2019quality,bouchaud2016excess,novy2013other} These quality stocks recovered very quickly from the crash, i.e., showing V-shaped recovery. On the other hand, many stressed stocks did not show signs of speedy recovery.~\cite{mahata2021characteristics,mahata2021modeling} 

Recently, the stock price dynamics during the COVID-19 pandemic is modeled by A. Mahata et al.~\cite{mahata2021modeling} They considered financial anti-fragility of the company and fund-flow by the foreign and domestic institutional investors as the main parameter and variable of the model, respectively. The parameter financial anti-fragility shows the financial strength of a company to fulfill its short term liabilities. It acts as a survival indicator of a company during a crisis. Hence, investors look for such financially antifragile stocks to get higher returns during a crisis.~\cite{mahata2021modeling,asness2019quality,bouchaud2016excess,novy2013other, castro2006integrated,lee2009corporate, bloomberg_lyz} Institutional investors also play an important role in the price movement as they are (a) responsible for huge fund inflow/outflow in the market; (b) they influence the smaller investors significantly.~\cite{mahata2021modeling,cao2008empirical, coval2007asset,ulku2014identifying,kling2008chinese,edelen2001aggregate,KP2020Dynamics,edelen2001aggregate,edelen1999investor,warther1995aggregate} However, in the sectors like banking, finance, realty and IT, the stocks and indices even with positive financial antifragility took longer time to recover during the COVID-19 pandemic, i.e., they displayed U-shaped and Swoosh-shaped recovery during the prolonged COVID-19 pandemic.~\cite{mazur2021covid} The reason for U-shaped recovery is that after the sharp crash, many factors like Government strict measures, relaxation in loan payment, closure of construction sites lead to a prolonged drop and late recovery. The IT sector showed a Swoosh-shaped recovery as after the sharp crash, the recovery was slow due to less inflow of funds. The study of the dynamics of these recoveries are important. Hence, the model proposed by A. Mahata et al.~\cite{mahata2021modeling} needs to be modified to describe such dynamics.

A study on investors' expectations following the stock market crash in March 2020 revealed that the investors largely became pessimistic about stock market returns in the short run (less than 1 year) even when their long run (10 years) expectations of the economy and stock markets remained largely unaltered.~\cite{giglio2021joint} A similar study on households' stock market outlook observed that individuals who have experienced loss during past stock market crashes expect slower recovery.~\cite{hanspal2020exposure} In response to the rapid spread of the COVID-19, strict lockdown measures were implemented in different countries. These factors lead to negative sentiment among the investors, that was reflected in the stock prices of various indices and companies.~\cite{ashraf2020economic,alexakis2021covid}  Generally, if the sentiment of the investors remain negative, recovery becomes slow. However, announcements of certain measures by the Governments helped to change the sentiment of the investors towards the stock market that lead to the recovery.~\cite{narayan2021covid}

In India, on 26\textsuperscript{th} March 2020, the finance minister(FM) announced the Pradhan Mantri Garib Kalyan Yojana of Rs. 1.7 trillion.~\cite{may12,kpmg,ghosh2020critique} On the following day, the Reserve Bank of India (RBI) reduced the repo rate by 75 basis points to 4.40 percent, lowered the cash reserve ratio by 100 basis points to 3.00 percent, and announced other measures that together aimed at injecting liquidity of Rs. 4.74 trillion into the system. On 12\textsuperscript{th} May 2020, the Prime Minister announced the comprehensive Atmanirbhar package for the Indian economy that worked out to 10 percent of the Indian GDP, making it among the largest in the world.~\cite{package} The package, after including the previous announcements by the FM and the RBI, promised for a economic stimulus of Rs. 20 trillion in multiple tranches. Given the unabated spread of the COVID-19, additional relief measures of Rs. 2.65 trillion was announced on 14\textsuperscript{th} November 2020. In addition to these relief measures, the decrease in the COVID cases and easing of the lockdown rules lead to the positive change in the investors sentiment that resulted in the recovery of the stock price. Hence, in order to simulate the stock price, the existing model needs to be modified to include the investor's sentiment.

The main aim of this paper is to propose an updated model incorporating investors' sentiment in the existing model formulated by A. Mahata et al.~\cite{mahata2021modeling} Our proposed model simulates the stock price movement of the sectoral indices that showed U-shaped and Swoosh-shaped recovery during the COVID-19 pandemic. We have considered the normalized net fund-flow due to institutional investors, sentiment and financial antifragility as the variables and parameter in the modified model, respectively. The simulation have been done for the real normalized fund-flow data ($\Psi_t$) and also for artificially generated fund-flow ($\Psi_{st}$). The simulation results are consistent with the real stock price movement during the pandemic. We have also identified the time-scale of shock and recovery of the sectoral indices. $\Psi_{st}$ is used to analyze the effects of different shock lengths, anti-fragility, negative sentiment lengths and fund-flow during recovery on the stock price movement. 

The rest of the paper is organized as follows: Sec.~\ref{sec:model} describes the model studied in this paper. Sec.~\ref{sec:moa} represents the technique used for the estimation of the time-scale. Sec.~\ref{sec:rd} represent the results obtained. Finally, we have concluded the results in Sec.~\ref{sec:con}.   

\section{Model formulation}
\label{sec:model}

A modified model for the stock price has been developed for the shock and recovery period during the COVID-19 pandemic. The Subsec.~\ref{EM} states the existing model and Subsec.~\ref{RM} describes the formulation of the modified model using the sentiment variable.

\subsection{Existing Model}
\label{EM}

A stock price model developed by A. Mahata et al.~\cite{mahata2021modeling} can explain V-shaped and L-shaped recovery of the stock market during the COVID-19 crisis. $\Psi_t$ by the foreign institutional investors and domestic institutional investors is considered as the main variable of the model. The basis for consideration of $\Psi_t$ as a variable is due to the facts that institutional investors are (a) responsible for huge fund inflow/outflow in the market; (b) the influence of institutional investors on the retail investors are significant.~\cite{mahata2021modeling,cao2008empirical, coval2007asset,ulku2014identifying,kling2008chinese,edelen2001aggregate,KP2020Dynamics,edelen2001aggregate,edelen1999investor,warther1995aggregate,barber2008retail} $\Psi_t$ is defined as the difference between the fund inflow and fund outflow by the Foreign Institutional Investor (FII) and Domestic Institutional Investor (DII). Let $\Delta D_t$ be the net fund-flow (inflow or outflow) due to FII and DII combined at time t.

\begin{equation}
\Psi_t= \frac{\Delta D_t}{max(abs(\Delta D_t))}
\label{eqn:psi}
\end{equation}
$\Psi_t$ is the normalized net fund-flow that is used for stock price updation.

An important parameter of the model is the financial antifragility ($\phi$) which is defined as the ratio of the difference between current assets ($\chi$) and current liabilities ($\zeta$) to operating expenses ($\xi$). $\phi$ signifies the company's ability to mitigate the current liabilities with their current assets. Current assets are those assets that can be converted in liquidity during the current operating cycle to meet up with the current liabilities. $\phi$ of $i^{th}$ company is calculated as

\begin{equation}
\displaystyle\phi_{i}=\frac{\chi_{i}-\zeta_{i}}{\xi_{i}}.
\label{eqn:phi1}
\end{equation}
 
The $\phi$ for a sectoral index can be written as 
\begin{equation}
\displaystyle\phi=\frac{\sum_{1}^{N} \phi_{i}}{N},
\label{eqn:phi2}
\end{equation}
where N is number of company in any sectoral index. $\phi$ acts as the control parameter for the price movement.

Finally, the model is defined as

\begin{eqnarray}
P_{t+1}&=&P_t\{1+\lambda\Psi_t\},~~~~~~during~shock\label{eqn:emodel1}\\
P_{t+1}&=&P_t\{1+\lambda\Psi_t\phi\},~~~~~otherwise
\label{eqn:emodel1}
\end{eqnarray} 	

During the shock period, the model simulation is independent of $\phi$ as the investors try to sell their holding irrespective of the fundamentals of the company to increase the cash holding. For Indian stocks, the value of $\phi$ is usually in the range $-2<\phi<2.$

$\lambda$ is the coefficient of $\Psi_t$ which shows the portion of net fund-flow towards a particular sector/company by the institutional investors. $\lambda$ is different for different sectors. Though this model explains the V- and L-shaped recoveries of stocks, it fails to explain the U- and Swoosh-shaped recoveries that depends mainly on the sentiment. Sentiment-based modification is discussed below.

\subsection{Modification of the existing Model: Sentiment based analysis}
\label{RM}

There are many stocks with positive anti-fragility in sectors like banking, finance, realty and IT that did not show V-shaped recovery during the COVID-19 pandemic. They rather remained in L-shaped for a long period. The main reason is that the sentiment regarding these sectors remained negative or neutral for a considerable period of time because of grim business outlook and strict lockdown measures. However, the sentiment turned positive due to various measures taken by the Government. In India, different measures like decrease in repo rate and cash reserve ratio by the RBI, economic stimulus packages by the Indian Government, decrease in the COVID cases and easing of lockdown measures helped to change the sentiment positively towards the stock market. The effect of such measures in stock market is observed during the second financial quarter of 2020. After that, the above sectors rallied for a prolonged period. Hence, sentiment of the investors is one important factor that plays an important role in the stock price movement.~\cite{guo2017can,baker2006investor}

The sentiment variable ($\theta$) can be negative sentiment, neutral sentiment and positive sentiment. When the investors are bearish towards the market or a particular sector, we assign a negative sentiment in the model. In a similar way, positive sentiment is assigned when the investors are bullish. A neutral sentiment describes a situation where the investors have no interest in the sector/company. Neutral sentiment is an ideal case as there will always be buying and selling of stocks. Mathematically, $\theta$ is defined as

\[
  \theta =
  \begin{cases}
            +1, & \textit{if sentiment is positive} \\
			       ~0, & \textit{if sentiment is neutral}\\
						-1, & \textit{if sentiment is negative} 
  \label{eqn:model1}
	\end{cases}
\]

So, the modified model can be written as 

\begin{eqnarray}
P_{t+1}&=&P_t\{1+\lambda\Psi_t\},~~~~~~during~shock\label{eqn:model2}\\
P_{t+1}&=&P_t\{1+\lambda\Psi_t\phi\theta\},~~~~~otherwise\label{eqn:model3}
\end{eqnarray}

where $\Psi_t$, $\phi$, $\lambda$ are the net normalized fund-flow by the institutional investors, financial anti-fragility and coefficient of $\Psi_t$, respectively which is discussed in detail in Subsec~\ref{EM}. The value of $\lambda$ is guessed on the basis of the real normalized net fund-flow due to mutual fund (MF) and foreign portfolio investor (FPI) which is shown in Fig.~\ref{fig:psi}. For Indian market, the time duration of negative $\theta$ is from 3 months to 7 months. 

In Subsec.~\ref{subsubsec:covid}, we have simulated the stock price movement from Eqn.~\ref{eqn:model2}-~\ref{eqn:model3} using $\Psi_{st}$ to study the dynamics of stock price during COVID like shocks. Further, in Subsec.~\ref{subsubsec:covid-19}, we have studied the stock price dynamics using the $\Psi_{t}$ and $\phi$ of the sectoral indices in Eqn.~\ref{eqn:model2}-~\ref{eqn:model3}.

\subsubsection{COVID-like shock}
\label{subsubsec:covid}

In order to understand the effect of COVID like crisis on the stock price, we need to study the model under different shock lengths ($T_S$), negative sentiment lengths ($T_N$), financial anti-fragility ($\phi$) and fund-flow during the recovery period. We have divided the fund-flow distribution into five segments, i.e., normal, shock, negative sentiment, recovery and post recovery periods. We have taken the distribution of $\Psi_{st}$ for those periods as $\mathcal{N}(0,0.28)$, $\mathcal{N}(-0.237,0.56)$, $\mathcal{N}(0.009,0.41)$, $\mathcal{N}(0.17,0.53)$, and $\mathcal{N}(0.051,0.43)$, respectively. The distribution is similar to the distribution of real fund-flow data. The values of $\lambda$ are chosen from Fig.~\ref{fig:psi} in ad-hoc basis depending on the portion of fund-flow in various sectors. In the present study, the values of $\lambda$ during normal, shock, negative sentiment, recovery and post recovery period are $\lambda=0.3,~0.4,~0.2,~0.6,~0.3$, respectively.

In order to understand the U-shaped and Swoosh-shaped recovery of stocks during different crisis, we have carried out the simulation considering different possibilities. Firstly, we have fixed $\phi=0.9$ and period of negative sentiment, $T_N=50$~day~(D), and varied $T_S=15$ D,~30~D,~45~D, and 60~D, respectively. Secondly, for fix $T_S=25$ D and $T_N=130$ D, $\phi$'s are varied as $\phi=0.6,~0.8,~1.0$ and $1.2$. Thirdly, we have varied the negative sentiment periods as $T_N=25$ D,~50~D,~75~D, and $100~D$ keeping $\phi=0.9$ and $T_S=25$ D. Finally, we have varied the fund-flow during the recovery period by changing the $\lambda$ value as $\lambda=0.05, 0.15, 0.25, 0.35$ keeping $\phi$, $T_S$ and $T_N$ at 0.9, 25 D and 0 D, respectively. The detailed analyses results are discussed in Subsec.~\ref{subsec:rd_covidlike}.

\subsubsection{COVID-19 shock}
\label{subsubsec:covid-19}

The model simulation of stock price during the COVID-19 has been done considering $\Psi_t$, $\phi$ and $\theta$ for the Indian stock market. $\theta$ represents the sentiment of the investors. $\Psi_t$ is the real net normalized fund-flow due to the institutional investors. The daily fund-flow data is available in the Moneycontrol website.~\cite{moneycontrol} $\phi$ of a company is estimated using Eqn.~\ref{eqn:phi1}. The current assets, current liabilities and operating expenses are taken from the financial report of a company that is available in Bombay Stock Exchange Ltd.~\cite{bse} 

\begin{figure}
\includegraphics[angle=0, width=7.5cm]{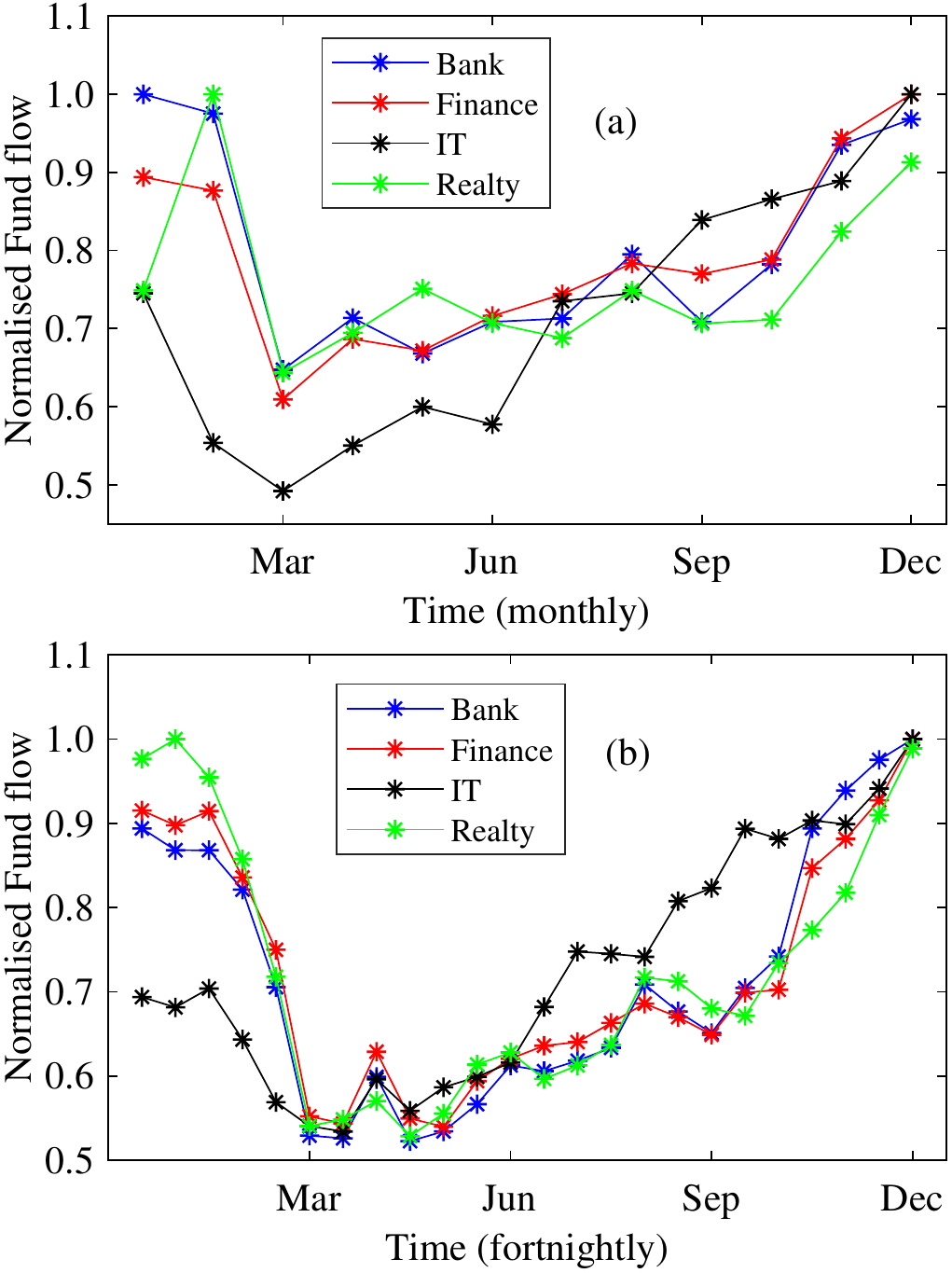}
\caption{\label{fig:psi}Plot (a) represents the monthly data of the normalized fund-flow due to mutual fund in the Bank, Finance, Realty and Information Technology (IT) sectors. Plot (b) represents the fortnightly data of the normalized fund-flow due to FPI in the Bank, Finance, Realty and Information Technology (IT) sectors.}
\end{figure}

The coefficient $\lambda$ is estimated from the fund-flow to a particular sector due to the DIIs and Foreign portfolio investors (FPIs). In this study, the sectoral fund-flow data due to Mutual fund investor (MFI) is taken as DII data. Fig.~\ref{fig:psi}(a) represents the MFI normalized monthly fund-flow in  Realty (Green), Information Technology (Black), Finance (Red) and Bank (Blue). These sectoral fund-flow data are obtained from Securities and Exchange Board of India (SEBI).~\cite{sebi} Fig.~\ref{fig:psi}(b) represents the FPI normalized fortnightly fund-flow in Realty (Green), Information Technology (Black), Finance (Red) and Bank (Blue). The data are taken from National Securities Depository Ltd. (NSDL), India.~\cite{nsdl} The plots clearly show that the fund-flow decreased drastically during the onset of the COVID-19 pandemic. It shows that after the crash, the infusion of funds was not rapid in the above sectors. This indicates that the investors are cautious to invest in these sectors. Further, the plots show that the fund-flow in the IT sector increased gradually compared to the other sectors. The inflow of funds may be due to the change in sentiment from bearish to bullish as compared to other sectors. The sectors like Bank, Finance and Realty took longer time to recover due to less inflow of funds. The main reason of less fund-flow in these sectors is negative sentiment. Considering the proportionality of fund-flow in various sectors, we have chosen the value of $\lambda$ as $\lambda=0.4,~0.9,~0.05,~0.4,~0.3$ for the Nifty Bank, $\lambda=0.4,~0.8,~0.05,~0.5,~0.4$ for the Nifty Financial and $\lambda=0.5,~0.9,~0.05,~0.4,~0.5$ for the Nifty Realty during pre-covid, shock, negative sentiment, recovery and post recovery period, respectively. As there was no negative sentiment towards the Nifty IT, we have chosen $\lambda=0.5,~0.6,~0.2,~0.8$ for pre-covid, shock, recovery and post recovery period, respectively. The results of the analysis is given in Subsec.~\ref{subsec:rd_covid}.

The negative sentiment in banking, finance, realty sectors during the COVID-19 pandemic can be confirmed by estimating the time scales of shock and recovery of these sectors. The time scale of shock and recovery of the IT sector is also calculated to confirm that it did not have a negative sentiment, but its recovery was not swift. Sec.~\ref{sec:moa} describes the technique used in the calculation of the time-scale of the above sectoral indices. The results of the time-scale analysis is given in Subsec.~\ref{subsec:timescale}.

\section{Techniques for  time-scale analysis}
\label{sec:moa}
In order to identify and analyze the different relevant time-scales ($\tau$) that our data-set inherits, we employ the Hilbert-Huang transform (HHT).~\cite{huang1998empirical} HHT consists of two parts; (a) Empirical mode decomposition (EMD) and (b) Hilbert spectrum. For a time-series that is either stationary or linear, its intrinsic time-scale can easily be computed. In the face of non-stationary and nonlinear behavior presented by the data, such an inference becomes non-trivial. It is evident that our stock-market data is far from being adherent to the assumptions of linearity and weak-stationarity; see~\cite{huang2003applications,huang1998empirical} for the assumptions. Empirical mode decomposition is one of the handful of successful data analysis tools available for such time-series data. In this subsection, we give a very brief overview of how the method has been designed to work, along with a statistical treatment to discern which intrinsic time-scales are particularly relevant and statistically significant.

\subsection{\label{subsec:emd} EMPIRICAL MODE DECOMPOSITION (EMD)}
As discussed in~\cite{huang1998empirical}, the EMD method uses the \emph{sifting process} to decompose an empirical time-series $X_t$ into simple oscillatory modes, each of them is called an \emph{intrinsic mode function} (IMF). 

More specifically, one can write
$$X_t=\sum_{n=1}^k C_n(t) + r(t),$$
where $C_n(t)$ is the $n$-th IMF and $r(t)$ is the residue which is usually a monotone function capturing the average trend of the data. Each IMF satisfies to following properties:
\begin{enumerate}
    \item In the entire data-set, the number of extrema (maxima and minima) and zero crossings should be equal or differ by one, and
    \item At any point, the mean value of the envelope defined by the local maxima and the envelope defined by the local minima should be zero.
\end{enumerate}
For more details on the sifting process and the mathematical properties of IMFs, the reader is encouraged to consult.~\cite{huang1998empirical}

Despite the analogy with simple oscillations, an IMF can have time-varying amplitude and frequency. The instantaneous frequency of any IMF, $C_n$, can be computed using its Hilbert transform:
$$D_n(t)=\frac{1}{\pi}PV\left(\int_{-\infty}^{+\infty}\frac{C_n(s)}{t-s}ds\right),$$ 
where $PV$ denotes the Cauchy principal value of the integral.
By the nature of Hilbert transform,
we can have an analytic signal,
$$Z_n(t)=C_n(t)+iD_n(t)=a_n(t)e^{i\theta_n(t)},$$
where $$a_n(t)=\sqrt{C_n^2(t)+D_n^2(t)}\mbox{ and } \tan{\theta(t)}=\frac{D_n(t)}{C_n(t)}.$$
The instantaneous frequency of the IMF is then calculated by:
$\omega(t)=\frac{d\theta(t)}{dt}$. To each IMF, we can also associate its time-scale $\tau$ by averaging $\frac{1}{\omega}$ over the entire data length. The time-scale of IMFs are particularly important to identify relevant events, such as shock, recovery, etc., and their persistence. It is worth noting that the sifting process picks out the IMFs in the increasing order of their time-scale. 

\subsection{\label{sec:s_test} STATISTICAL SIGNIFICANCE TEST (SST)}
Our analysis also entails a statistical treatment of the data-set, assuming that our data might be erroneous. As with any physical signal, the true stock-price data is also inevitably mixed with different types of noise.
The non-linear and non-stationary nature of the data makes it difficult to filter the noise from the true signal. However, we can assign statistical significance to the computed IMFs to decide whether they are just white noise, or they really carry some information/trend.~\cite{wu2004study,wu2005statistical} To build context for comparison, we follow~\cite{wu2004study} to briefly discuss the probability distribution of the IMFs and their energy density, when computed from white noise.

The energy density $E_n$ of the $n$-th IMF, $C_n$, is defined by 
$$E_n = \frac{1}{N}\sum_{j=1}^N C_n^2(j),$$ where $N$ denotes the number of data points. We also define the notion of mean period $T_n$ of $C_n$ to be the number of local maxima and diving it by $T$, the total length. As empirically observed in~\cite{wu2004study}, the IMFs (except for $n=1$) from a normalized Gaussian white noise data-set have the identical spectral shape in terms of the $\ln{T}$-axis. The observation leads to the the following relation for the $n$-th IMF ($n\neq1$) of a Gaussian white noise series:
\begin{equation}\label{eqn:ET}
\ln{\overline E_n} + \ln{\overline T_n} = 0,
\end{equation}
where 
$$
\overline T_n = \frac{\int S_{\ln{T},n} d[\ln{T}]}{\int S_{\ln{T},n} d[\ln{T}]/T}
$$
is the spectrum-weighted mean period and $\overline E_n$ is the mean energy density. For very large $N$, the $E_n$ and $T_n$ are good estimators of $\overline E_n$ and $\overline T_n$, respectively. 

Using the equation \ref{eqn:ET}, the probability distribution of the energy density of the IMFs of a Gaussian white noise data-set is also established. The the distribution of $y=\ln{E_n}$ has been shown to approximately follow $\mathcal{N}(-x,\frac{2}{N}e^{x})$, where $x=\ln{\overline T_n}$. As a consequence, one can derive the energy spread lines at
different level of significance using the following formula:
$$
y = -x \pm k\sqrt{\frac{2}{N}}e^{\frac{x}{2}},
$$
where $k=2.326$ a for $99\%$ confidence interval as an example. The $(T_n,E_n)$ pair for the $n$-th IMF $C_n$ that contains information is expected to lie outside the spread lines at $99\%$ confidence level.~\cite{wu2004study}

\section{Results and discussion}
\label{sec:rd}
We have presented the simulation result of COVID-like shock using $\Psi_{st}$ in Subsec.~\ref{subsec:rd_covidlike}. In Subsec.~\ref{subsec:rd_covid}, we analyze the simulated stock price obtained using $\Psi_{t}$ during the COVID-19 pandemic for the (1) Nifty Bank, (2) Nifty Financial, (3) Nifty Realty and (4) Nifty IT. The result of time-scale analysis of these indices is presented in Subsec.~\ref{subsec:timescale}. The daily stock price data are downloaded for the period of $1^{st}$ July 2019 to $31^{st}$ May 2021 from NSE website.~\cite{nse}

\subsection{Simulation of COVID-like shock}
\label{subsec:rd_covidlike}
\begin{figure}
\includegraphics[angle=0, width=8.5cm]{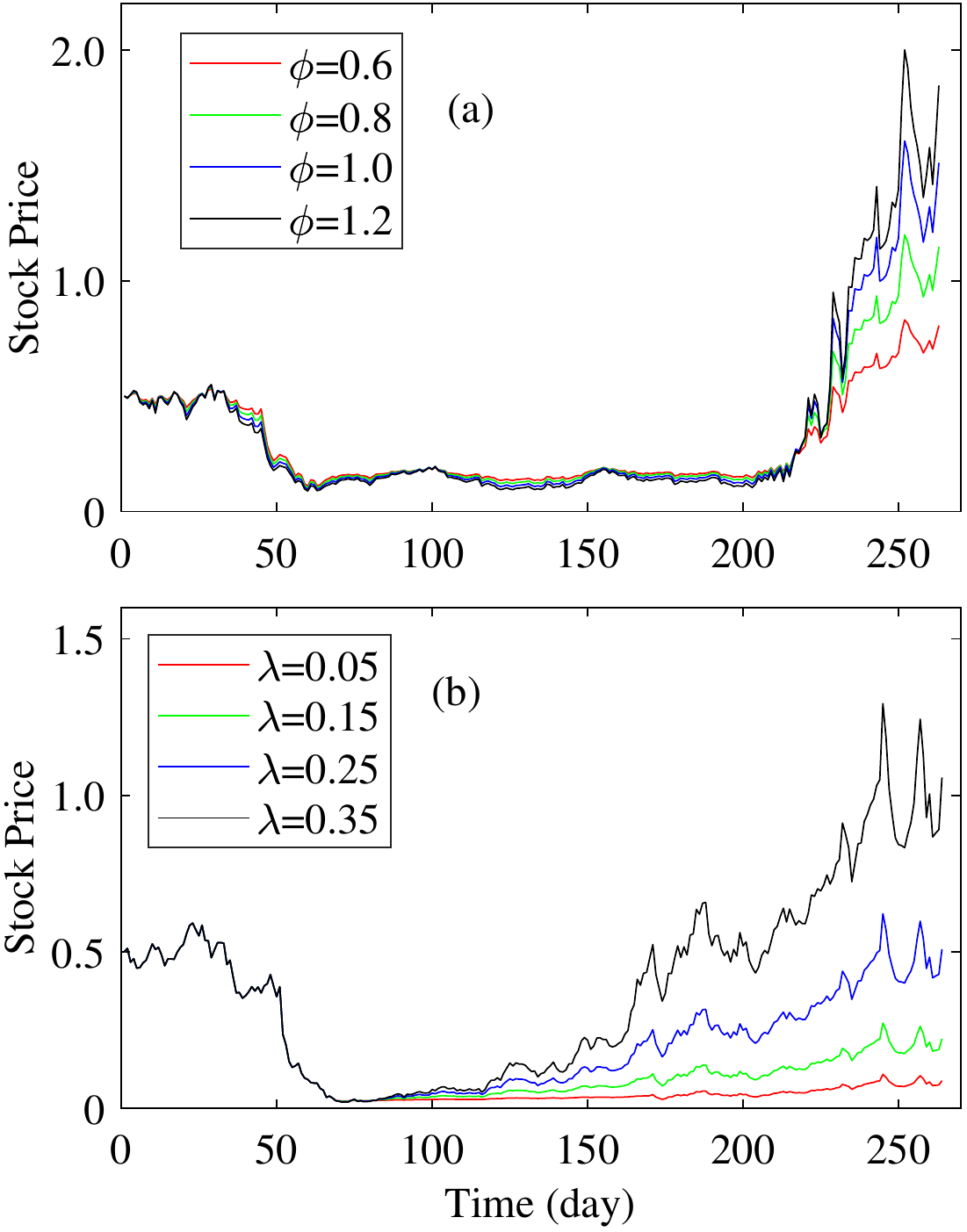}
\caption{\label{fig:SD}Plot (a) shows the U-shaped recovery using synthetic data for fixed $T_S=25~D$ and $\phi=0.6,~0.8,~1.0$ and 1.2. Plot (b) represents the Swoosh-shaped recovery for fixed $\phi=0.9$, $T_S=25~D$ and $\lambda=0.05, 0.15, 0.25, 0.35$ during the recovery period. $D$ represents day and $\phi$ represents the financial antifragility. For the simulation, initial condition is taken as 0.5}
\end{figure}

Fig.~\ref{fig:SD}(a) represents the plot of U-shaped recovery of stock price for different $\phi$ using $\Psi_{st}$ for a fix $T_S$=25~D and $T_N$=130~D. The plots show the simulated stock price for $\phi= 0.6, 0.8, 1.0, 1.2$ in red, green, blue and black color, respectively. During the negative sentiment period, they did not show any signs of recovery even though $\phi >0$. When the sentiment becomes positive, the stocks with positive $\phi$ starts recovering. The figure shows that the stocks with the highest $\phi$ recovers rapidly and outperforms its peer. The rate of recovery decreases with decrease in the value of $\phi$. These analyses show that the recovery rate directly depends on the $\phi$ when the sentiment is positive.

In order to understand the Swoosh-shaped recovery, we have considered $T_N=0~D$, $T_S=25~D$ and $\phi=0.9$. Fig.~\ref{fig:SD}(b) represents the Swoosh-shaped recovery for different portion of fund allocated ($\lambda$) during the recovery period. The plots in red, green, blue and black represent the stock price for $\lambda$ as 0.05, 0.15, 0.25 and 0.35 during the recovery period, respectively. The figure shows that when there is no negative sentiment, price starts recovering after the shock period is over. The recovery depends only on the fund-flow by the investors. Fig.~\ref{fig:SD}(b) shows that the stock with highest $\lambda$, i.e., 0.35 recovers rapidly. The recovery of stocks with $\lambda= 0.05, 0.15$ is slow. The rate of recovery is directly proportional to the fund-flow. Hence, the recovery not only depends on the sentiment but is also dependent on the portion of fund allocated.

\begin{figure}
\includegraphics[width=8.5cm]{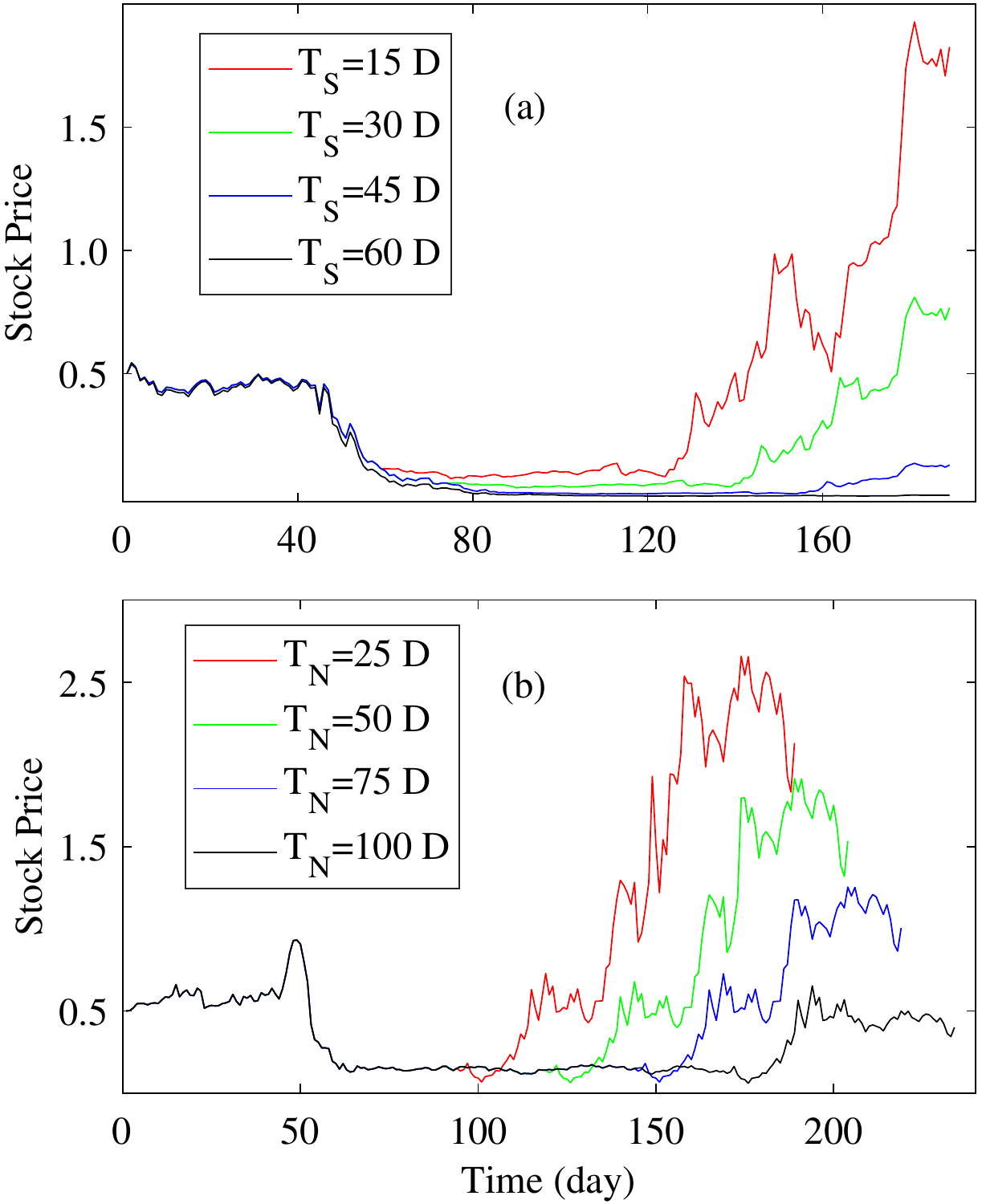}
\caption{\label{fig:SD_Senti}Plot (a) represents the U-shaped recovery using the synthetic data for fixed $\phi=0.9$ and $T_S=15~D,~30~D,~45~D$ and $60~D$. Plot (b) represents the U-shaped recovery using synthetic fund-flow data for fixed $\phi=0.9$ and $T_S=25~D$ with $T_N=25~D,~50~D,~75~D$ and $100~D$. For the simulation, initial condition is taken as 0.5. $D$ and $\phi$ represent day and financial antifragility.}
\end{figure}

Fig.~\ref{fig:SD_Senti}(a) represents the U-shaped recovery for different $T_S$ with $\phi=0.9$ and $T_N=50~D$. The plots show the simulated stock price for $T_S=15~D, 30~D, 45~D, 60~D$ in red, green, blue and black, respectively. The figure shows that the stocks with shortest shock length, i.e., $T_S=~15~D$ recovers quickly to its initial level after the sentiment turns positive. An extended $T_S$ will also affect the recovery period. The stock with the least $T_S$ will have maximum recovery period. The recovery period will decrease as the $T_S$ increases. This is clearly seen in Fig.~\ref{fig:SD_Senti}(a) for stocks with $T_S=45~D, 60~D.$ Hence, the recovery of positive $\phi$ stocks also gets affected with the increase in $T_S.$

Fig.~\ref{fig:SD_Senti}(b) represents the U-shaped recovery for different $T_N$ with $\phi=0.9$ and $T_S=25~D$. The plots show the simulated stock price for $T_N=25~D, 50~D, 75~D, 100~D$ in red, green, blue and black, respectively. The figure shows that for a fix $\phi$, the recovery of stocks depend on the $T_N$. Fig.~\ref{fig:SD_Senti}(b) shows that the stocks with $T_N=25~D$ shows faster recovery than the stocks with $T_N=50~D, 75~D, 100~D$. The recovery starts only after the sentiment becomes positive. A long $T_N$ affects the recovery period which can be seen in Fig.~\ref{fig:SD_Senti}(b). The recovery period decreases with an increase in $T_N$. It clearly shows that the sentiment of the investors is very important for the stocks to recover. Until the sentiment becomes positive, no fundamentally strong company can recover and performs well.

\subsection{Simulation of the COVID-19 shock}
\label{subsec:rd_covid}

\begin{figure}
\includegraphics[angle=0, width=8.5cm]{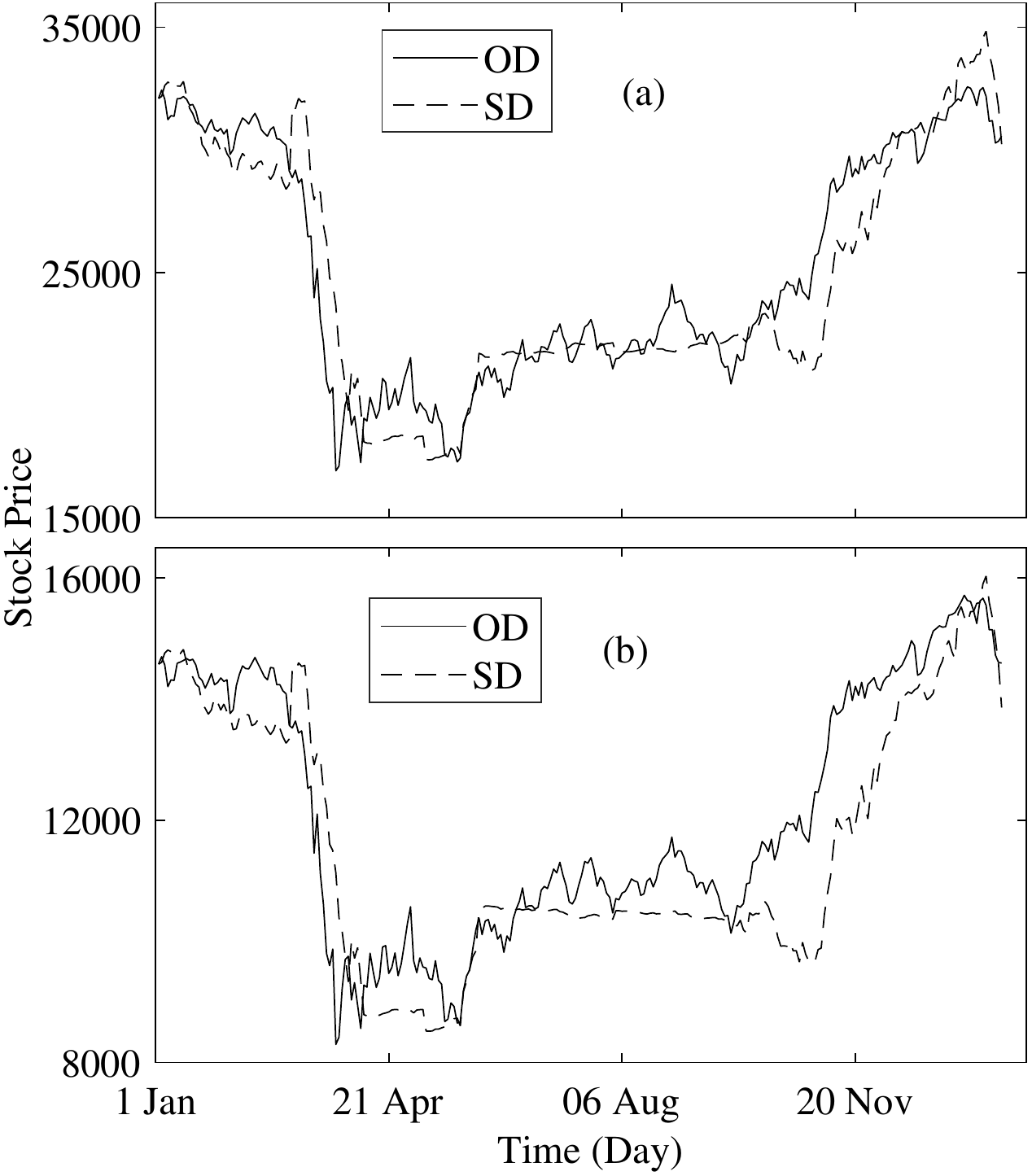}
\caption{\label{fig:OD_Simu1}Plot (a) represents the original stock price movement of the Nifty Bank ($-OD$) and its corresponding model simulated stock price movement ($--SD$) with $\phi=1.04 $ and $T_S=21~D$. Plot (b) represents the original stock price movement of Nifty  Financials ($-OD$) and its corresponding model simulated stock price movement ($--SD$) with $\phi=0.80$ and $T_S=21~D$}
\end{figure}

Figs.~\ref{fig:OD_Simu1}(a) and \ref{fig:OD_Simu1}(b) represent the model simulated stock price ($--$ line) using $\Psi_t$ and original stock price (solid line) of the Nifty Bank and Nifty Financial, respectively during the COVID-19 pandemic. The simulation results show that the stock price crash started from the $1^{st}$ week of March 2020 and fell for around 20-25~D till $1^{st}$ week of April 2020. After a severe crash, they did not show the signs of recovery for a considerable period as the investors had a very bearish/negative sentiment towards these sectors. Hence, these stocks could not recover though they were financially anti-fragile. In order the give a boost to reviving the economy, the Indian government announced a COVID relief package of Rs. 20 trillion on May 12, 2020~\citep{may12} This amount was to be invested in different sectors in different tranches including a special liquidity facility for Mutual Funds for Rs. 50,000 crore. The stock market's sentiment turned bullish, however, this sentiment did not last for long in the banking and financial sectors. Though, a slight recovery during the month of May is seen in these sectors due to the change of sentiment of the investors which is captured in both Original and Simulated stock price as shown in Fig.~\ref{fig:OD_Simu1}(a) and \ref{fig:OD_Simu1}(b). From the $4^{th}$ week of September 2020, strong recoveries in these sectors were seen due to the positive sentiment of the investors. During this period, the fund inflow towards these sectors was increased significantly as shown in Fig.~\ref{fig:psi}. The inflow of funds may also be due to the expectation of the investors that the government might announce some stimulus package for the recovery of economy. The government's announcement of a stimulus package of Rs. 2.65 lakh crore during the $2^{nd}$ week of November 2020~\citep{kpmg} and the decrease in the number of COVID cases made the investors even more bullish. As a result, there was regular inflow of funds which lead to the recovery of these sectors. The model simulation shows a U-shaped recovery which is consistent with the original stock price movement of the Nifty Bank and Nifty Financial. It may be due to the negative sentiment of the investors that led to these financially strong sectors for U-shaped recovery. 
 
\begin{table}
\footnotesize
\caption{ The table shows the correlation coefficient ($\nu$) and the p-value between the Original data (OD) and the simulated data (SD).}
\label{tab:1}
\begin{tabular}{|r|r|r|r|r|}
\hline
&Nifty Bank & Nifty Financial & Nifty Realty & Nifty IT \\
\hline
P-val & $9.86\times 10^{-120}$ & $1.41\times 10^{-105}$ & $1.81\times 10^{-76}$ & $1.44\times 10^{-136}$ \\
\hline
$\nu$ & 0.930 & 0.911 & 0.848 & 0.948 \\
\hline
\end{tabular}
\end{table}

\begin{figure}
\includegraphics[angle=0, width=8.5cm]{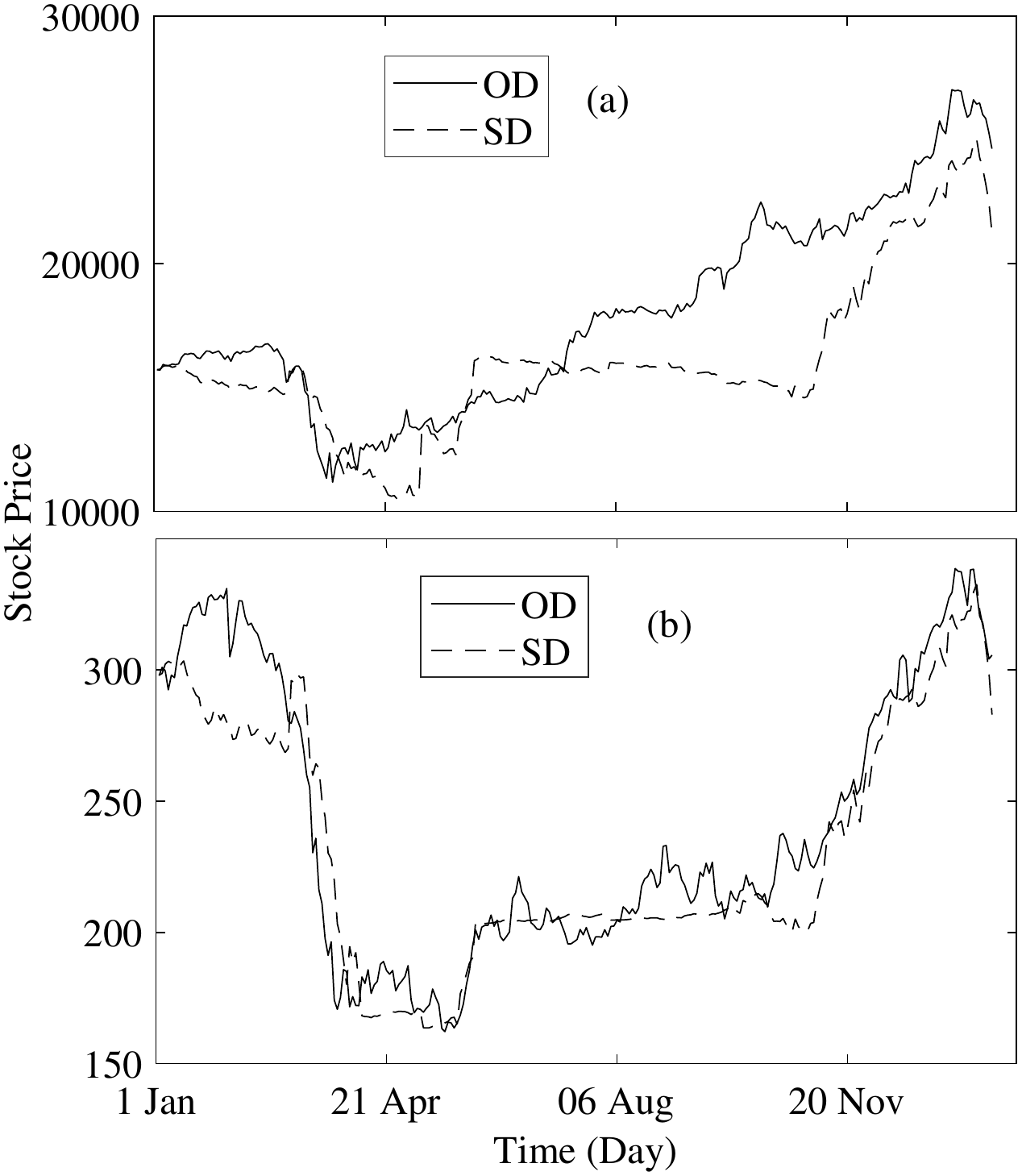}
\caption{\label{fig:OD_Simu2}Plot (a) shows the original stock price movement of the Nifty IT($-OD$) and its corresponding model simulated stock price movement ($--SD$) with $\phi=0.45 $ and $T_S=21~D$. Similarly, plot (b) shows the original stock price movement of Nifty Realty ($-OD$) and its corresponding model simulated stock price movement ($--SD$) with $\phi=0.71$ and $T_S=21~D$.}
\end{figure}

Figs.~\ref{fig:OD_Simu2}(a) and \ref{fig:OD_Simu2}(b) represent the original stock price (solid line) and simulated price ($--$line) for the Nifty IT and Nifty Realty, respectively. Fig.~\ref{fig:OD_Simu2}(a) shows that after the shock period, the IT sector recovers slowly showing Swoosh-shaped recovery. Though there was no negative sentiment towards IT sectors, the fund-flow in this sector was less compared to the Pharma and FMCG sectors.~\cite{mahata2021modeling} As a result, we have seen slow recovery in the IT sector. Fig.~\ref{fig:OD_Simu2}(a) shows that there is some deviation between simulated and original stock price during the months of August-September 2020. During this period, $\psi_t$ was negative though the fund-flow towards the IT sector was positive [Fig.~\ref{fig:psi}]. Simulation using $\Psi_t$ may have created such deviation. Similar to the Nifty Bank and Nifty Financial, the Nifty Realty also shows U-shaped recovery as seen in Fig.~\ref{fig:OD_Simu2}(b). Though the stocks with U-shaped showed some initial recovery during the month of May 2020 due to the announcement of COVID relief package, the sentiment of investors did not last long. Hence, their recovery did not last long as seen in Fig.~\ref{fig:OD_Simu2}(b). A steady recovery was seen from the month of November 2020 due to the stimulus package announcement by the Indian government and decrease in the COVID cases in India. These reasons lead to the change of the sentiment which resulted in the steady recovery of stock price.

We have also calculated the correlation coefficient ($\nu$) between original stock price and the simulated stock price which is shown in Table~\ref{tab:1}. $\nu$ values show that there is a high correlation between the real data and our model simulated data for all the four sectors. The p-value is also calculated for the same which is within a 95\% confidence level, i.e., less than 0.05.

\begin{table*}
%\footnotesize
%\tiny
\caption{Measure of events correlation coefficient ($\nu$) between original data (OD) and its IMFs. Significance test of the correlation coefficient of all the IMFs with the OD have been performed. $IMF_6$ shows the maximum correlation with the OD for Nifty Bank, Nifty Financial and Nifty Realty, whereas, in case of Nifty IT $IMF_5$ gives maximum correlation. The p-value of the $IMFs$ are also calculated for all the four indices. $\tau$ represents the time-scale of the shock and recovery in days (D)} 

\label{tab:2}
\begin{tabular}{|l|r|r|r|r|r|r|r|r|r|r|r|r|}
\hline
IMF & \multicolumn{3}{c|}{NIFTY BANK} & \multicolumn{3}{c|}{NIFTY FINANCIAL}& \multicolumn{3}{c|}{NIFTY REALTY}& \multicolumn{3}{c|}{NIFTY IT} \\
\cline{2-13}

& $\tau$  & P-val & $\nu~~~$ & $\tau$ & P-val & $\nu~~~$ & $\tau$  & P-val&  $\nu~~~$& $\tau$ & P-val & $\nu~~~$ \\
\hline
IMF1 & 4 & 0.095 & 0.102 & 4 & 0.189 & 0.079 & 4 & 0.503 & 0.040 & 4 & 0.462 & 0.057  \\
 \hline
IMF2 & 10 & $1.71\times 10^{-07}$ & 0.311 & 9 & 0.253 & 0.0696 & 6 & 0.712 & 0.023 & 8 & 0.525 & 0.0488  \\
 \hline
IMF3 & 20 & 0.011 & 0.153 & 18 & 0.159 & 0.086 & 14 & 0.741 & 0.020 & 20 & 0.990 & -0.001  \\
 \hline
IMF4 & 48 & 0.020 & 0.141 & 43 & $3.45\times 10^{-04}$ & 0.215 & 32 & 0.086 & 0.104 & 46 & 0.041 & 0.156  \\
 \hline
IMF5 & 94 & 0.012 & 0.152 & 121 & 0.003 & 0.177 & 80 & 0.166 & 0.084 & 142 & $3.60\times 10^{-40}$ & 0.804  \\
 \hline
IMF6 & 245 & $1.01\times 10^{-106}$ & 0.912 & 245 & $9.64\times 10^{-112}$ & 0.919 & 241 & $5.43\times 10^{-67}$ & 0.819 &  &  &   \\
 \hline

\end{tabular}
\end{table*}

\begin{figure}
\includegraphics[angle=0, width=8.5cm]{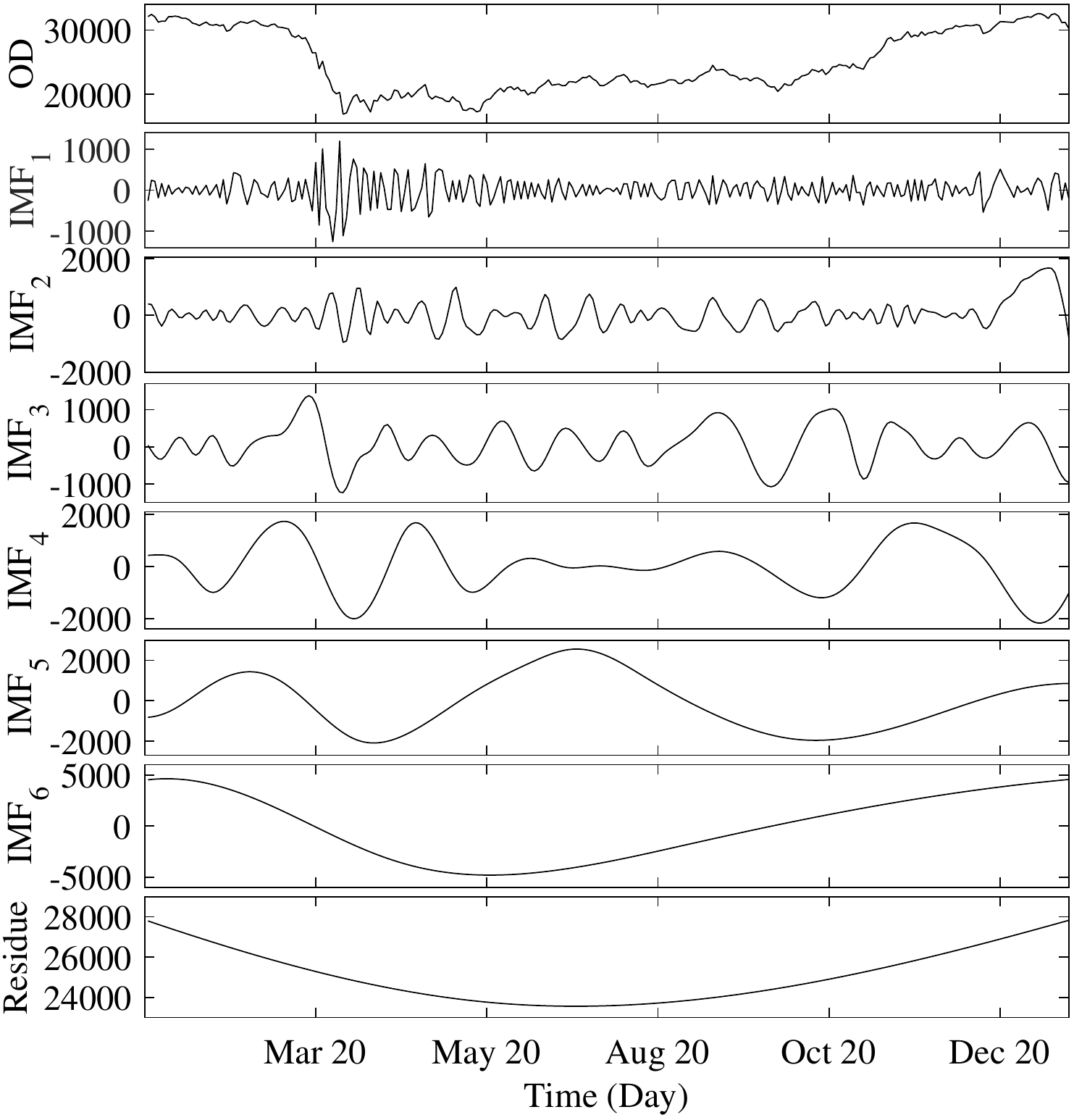}
\caption{\label{fig:IMF_plot}Plot represents the original data (OD) of the Nifty Bank and its $IMF_1$-$IMF_6$. Residue represents the overall trend of the original data.}
\end{figure}

\begin{figure}
\includegraphics[angle=0, width=8.5cm]{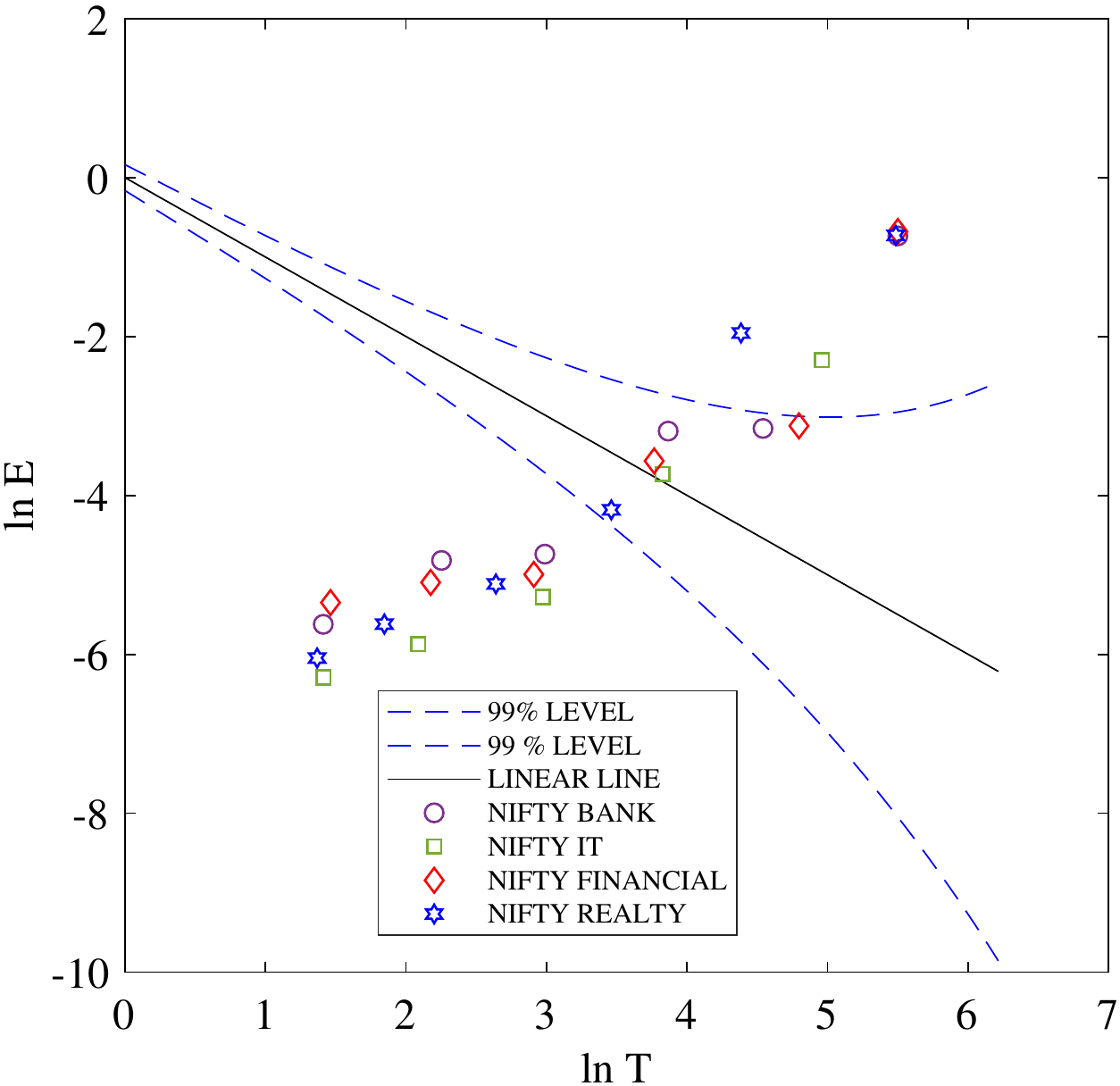}
\caption{\label{fig:Sig_Test}Plot represents the significance test of the IMFs of the Nifty Bank, Nifty Financials, Nifty Realty and Nifty IT sectoral indices with 99\% significance level.}
\end{figure}

\begin{figure*}
\includegraphics[angle=0, width=15cm]{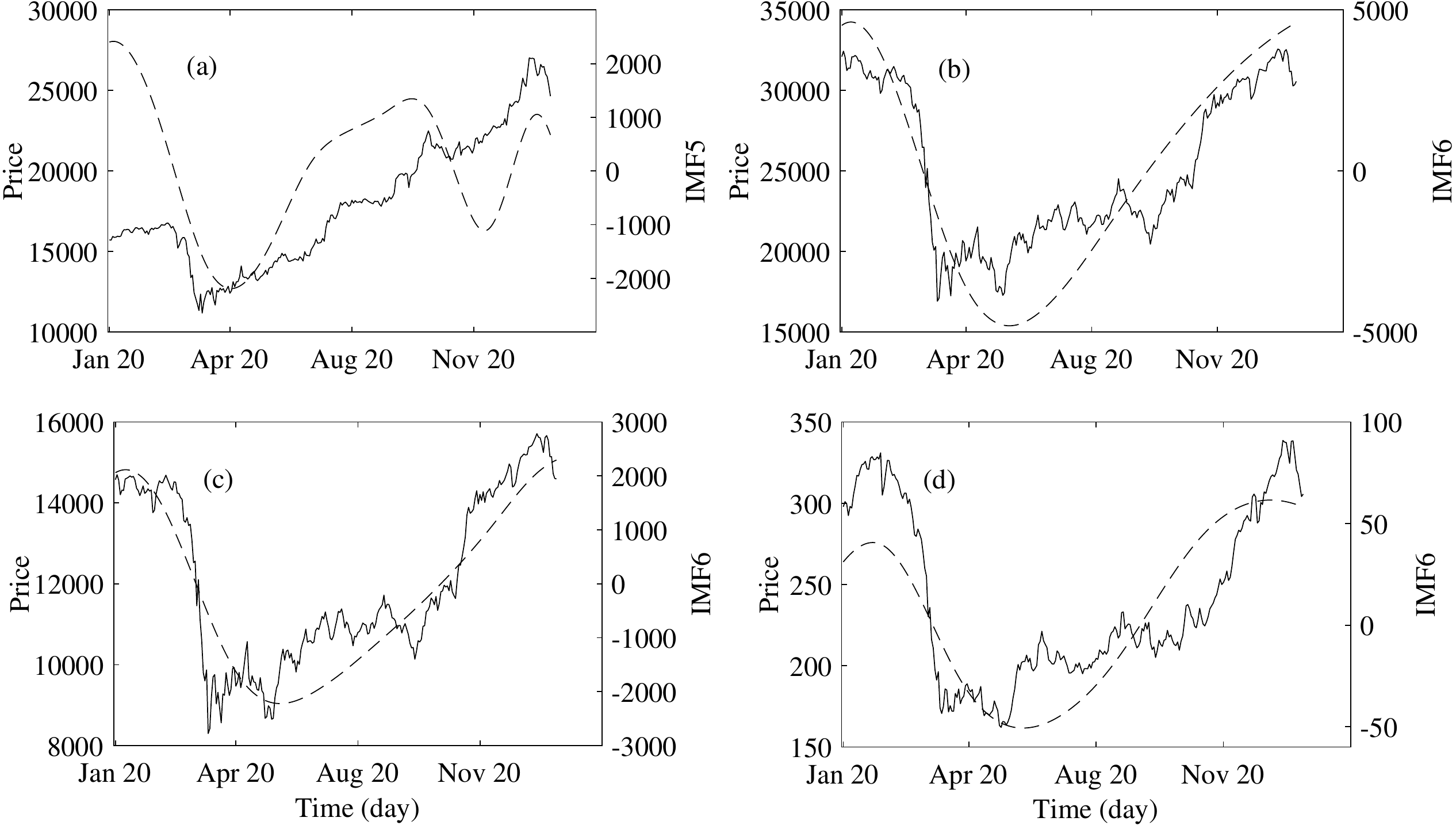}
\caption{\label{fig:Sig_IMF}Plot (a), (b), (c) and (d) represent the original data (Solid line) and its dominant IMF ($--$ line) of the Nifty IT, Nifty Bank, Nifty Financials and Nifty Realty, respectively.}
\end{figure*}

\subsection{Time-scale Analysis}
\label{subsec:timescale}
HHT technique is applied to calculate the time-scale ($\tau$) of shock and recovery of sectoral indices during the COVID-19 pandemic. Fig.~\ref{fig:IMF_plot} represents the original data (OD) of the Nifty Bank, its six IMFs and residue obtained from the EMD technique. The residue represents the overall trend of the OD. A statistical significance test (SST) of the decomposed IMFs of all the indices are carried out and are represented in Fig.~\ref{fig:Sig_Test}. The plot shows the SST at 99\% significance level. IMFs which lie within the significance level may contain noise. In case of Nifty Bank and Nifty Financial all the IMFs contain information except IMF4 and IMF5, i.e., IMF4 and IMF5 are inside the significance level. For Nifty IT, IMF4 lies within the significance level and for Nifty Realty, IMF5 is within the confidence level.

We have calculated the $\tau$, i.e., time-scale from shock till recovery, of U-shaped and Swoosh-shaped sectoral indices during the COVID-19 by identifying the dominant IMF. For this, we have calculated the correlation coefficient ($\nu$) and p-value between the original stock price and all the IMFs. Table~\ref{tab:2} contains the $\tau$, $\nu$ and p-value of all the IMFs. Considering this, and the result of the SST we have identified the dominant IMF. Figs.~\ref{fig:Sig_IMF}(a)-~\ref{fig:Sig_IMF}(d) represent the original data (solid line) and their dominant IMF ($--$ line) of Nifty IT, Nifty Bank, Nifty Financial and Nifty Realty, respectively. In case of the Nifty Bank, Nifty Financial and Nifty Realty, IMF6 is the dominant IMF and for the Nifty IT, IMF5 is dominant. The estimated $\tau$ for Nifty IT, Nifty Bank, Nifty Financial and Nifty Realty are 142 D, 244 D, 244 D and 241 D, respectively. The obtained $\tau$'s are consistent with the real time taken by the indices to recover during the COVID-19 pandemic.

\section{conclusion}
\label{sec:con}
In this paper, we have modified an existing model of stock price dynamics to explain the U-shaped and Swoosh-shaped recovery during the COVID-19 pandemic. In our simulation, we have considered net fund-flow by the institutional investors ($\Psi_{t}$) as the main variable, and financially anti-fragility ($\phi$) and sentiment of the investors ($\theta$) as the main parameters. In our simulation, we have considered that stock price crash is totally governed by $\Psi_{t}$. Whereas, $\theta$ and $\phi$  are important parameters during the recovery of stock price. The model simulates the stock price considering different $T_S$, $T_N$, $\phi$ and $\lambda$ during the recovery period. The model simulated result and the real stock price are also consistent during the COVID-19 pandemic. We have also identified the $\tau$ from the original data using the HHT technique.  

We have obtained U-shaped recovery of stocks for $\phi=0.6, 0.8, 1.0, 1.2$ using the synthetic fund-flow data ($\Psi_{st}$) with $T_N=130~D$ and $T_S=25~D$. After the sentiment of the investor changes to positive, the stock with a higher $\phi$ outperforms the others. We have also obtained U-shaped recovery of stocks with $\phi>0$ for different $T_S$. When $T_S=15~D$ and $30~D$, the stock price recovers to its pre-shock value after the $T_N$ is over. However, with an extended $T_S$ the recovery of positive $\phi$ stocks also becomes difficult. The recovery of stock under different $T_N$ is also studied. We have obtained U-shaped recovery for different $T_N$ with $T_S=25~D$ and $\phi=0.9$. The stock price recovers for $T_N=25~D$ and $50~D$ showing U-shaped recovery. When the $T_N$ is extended beyond 50~D the recovery of positive $\phi$ stocks also becomes difficult. Lastly, we have obtained Swoosh-shaped recovery of stocks for different $\lambda$ values during the recovery period with $T_N=0~D$, $T_S=25~D$ and $\phi=0.9$. When $\lambda= 0.25~\&~0.35$ the stock price recovers to its pre-COVID price showing Swoosh-shaped recovery. However, for $\lambda= 0.05~\&~0.15$ the recovery is very slow. With the decrease in $\lambda$ value during recovery, the rate of recovery also decreases. 

We have obtained U-shaped and Swoosh-shaped recovery of stock price from the model simulation using the real normalized net fund-flow data ($\Psi_t$). We have calculated the $\phi$ for Bank, Financial, Realty and IT sectoral indices. The Nifty Realty, Nifty Bank, Nifty Financial show U-shaped recovery from our simulation result and  Nifty IT shows a Swoosh-shaped recovery. We observe that the model simulation results are consistent with the real stock price movement. Finally, we have estimated the $\tau$ for the above indices and they are consistent with the real time taken by these indices to recover during the COVID-19 pandemic. The identification of U- and Swoosh-shaped stocks will help the investors plan their entry and exit positions during a crisis.

\section*{Acknowledgment}
We would like to thank Salam Rabindrajit Luwang, Sandeep Parajuli and Om Prakash for their valuable comments during the preparation of the manuscript. NIT Sikkim is appreciated for allocating doctoral research fellowships to A.R. and A.M.

\section*{Data availability}
The data used in this study are openly available in moneycontrol~\cite{moneycontrol}, NSDL~\cite{nsdl}, SEBI~\cite{sebi} and NSE India~\cite{nse}.
%\pagebreak
\bibliography{apssamp2}
\end{document}